\documentclass[a4paper,10pt]{article}

\usepackage{amssymb}
\usepackage{amsmath}
\usepackage[usenames,dvipsnames]{color}
\usepackage{hyperref}
\usepackage[left=.8in,right=.8in,top=.8in,bottom=.8in]{geometry}                  

\hypersetup{
    colorlinks=true,         
    linkcolor=blue,          
    citecolor=red,        
    urlcolor=Violet             
}

\newtheorem{theo}{Theorem}

\newcommand{\diby}[2]{\ensuremath{\frac{\delta #1}{\delta #2}}}

\def\be{\begin{equation}}
\def\ee{\end{equation}}
\def\bea{\begin{eqnarray}}
\def\eea{\end{eqnarray}}

\title{A first look at Weyl anomalies in shape dynamics}
\author{\bf Henrique Gomes\footnote{\href{mailto:gomes.ha@gmail.com}{gomes.ha@gmail.com}}\\\it  Department of Physics,  University of California, Davis,   CA, 95616}

\begin{document}

\maketitle

\begin{abstract}
One of the more popular objections towards shape dynamics is the suspicion that anomalies in the   spatial Weyl symmetry will arise upon quantization.  The purpose of this short paper is to establish the tools required for an investigation of the sort of anomalies that can possibly arise. The first step is to adapt to our setting Barnich and Henneaux's  formulation of gauge cohomology in the Hamiltonian setting, which serve to  decompose the anomaly into a spatial component and time component. The spatial part of the anomaly, i.e. the anomaly in the symmetry algebra itself ($[\Omega, \Omega]\propto \hbar$ instead of vanishing) is given by a projection of the second ghost cohomology of the Hamiltonian BRST differential associated to $\Omega$, modulo spatial derivatives.   The temporal part, $[\Omega, H]\propto\hbar$ is given by a different projection of the first ghost cohomology  and an extra piece  arising from a solution to a functional differential equation. Assuming locality of the gauge cohomology groups involved, this part is always local. Assuming locality for the gauge cohomology groups, using Barnich and Henneaux's results, the classification of Weyl cohomology for higher ghost numbers performed by Boulanger, and following the descent equations, we find a complete characterizations of anomalies in 3+1 dimensions. The spatial part of the anomaly  and the first component of the temporal anomaly are always local given these assumptions even in shape dynamics. The part emerging from the solution of the functional differential equations explicitly involves the shape dynamics Hamiltonian, and thus might be non-local.  If one restricts this extra piece  of the temporal anomaly to be also local, then overall no \emph{local} Weyl anomalies, either temporal or spatial, emerge in the 3+1 case.
 \end{abstract}

\section{Introduction}

One of the issues which complicates the introduction of Weyl symmetries in field theories in general is that these types of symmetries usually present anomalous behavior upon quantization \cite{Bertlmann}. Anomalies arise when a classical symmetry fails to be realized in the quantum theory.
This failure can be seen from a multitude of different perspectives: it is the failure of the measure of the path integral to transform covariantly even if one chooses a covariant regularization procedure \cite{Fujikawa}, or the failure of regularization counter-terms to obey the symmetries of the bare action.  Local gauge anomalies, unlike global anomalies, \cite{Adler, Bell-Jackiw} will in general spoil  renormalization of the quantum theory. 

{ Cohomological methods in field space are able to pinpoint which are the candidates for a given anomaly. However, they cannot yield their normalization, for which one would have to resort to explicit calculations of the quantized theory.} In the present paper we will \emph{not} study a calculation of e.g. the 1-loop effective action of shape dynamics, which would settle the issue of anomalies completely. Instead, we will employ the classical cohomological methods that can determine the exact functional form of possible anomalous terms, but not their normalization \cite{Bertlmann}.

Shape Dynamics is a novel formulation of gravity,  formulated in the Hamiltonian 3+1 formalism, in which refoliation invariance is replaced with local spatial conformal (Weyl) invariance. This formulation was motivated by Barbour's interpretation of Mach's principle and the construction of Shape Dynamics utilizes many results of the conformal approach to the initial value problem of ADM.  It possesses two dynamical propagating degrees of freedom, and has as kinematical variables the same $g_{ab}$ and $\pi^{ab}$ as the Hamiltonian version of ADM General Relativity. What is noteworthy about this model is the fact that it possesses spatial Weyl invariance, acting on both the metric and on the momenta. It maintains the correct number of degrees of freedom since it does not have refoliation invariance.    It is a theory that takes as its geometric observables \emph{spatial} conformal--diffeomorphism invariants, as opposed to \emph{space-time} diffeomorphism invariants.

Shape dynamics indeed possesses a type of ``hypersurface--intrinsic" phase space Weyl symmetry, thus a natural question to pose is the possibility that it too will develop Weyl anomalies. 
However, there is an added difficulty in the very formulation of the cohomological study of anomalies for Shape Dynamics, since it is an inherently Hamiltonian theory, which has not so far been formulated as a Lagrangian picture, much less a space-time one. This obstruction requires the use of some non-standard techniques for the cohomological computation itself, which fortunately were developed in the couple of papers by Barnich et al \cite{Barnich_Hamiltonian, Barnich&Henneaux}. We feel that shape dynamics, if nothing else, provides an interesting playground for the application of these mathematical techniques,  because the more standard cohomological treatment of anomalies are unsuited for the shape dynamics setting.

As a last point in this introduction, let us point out that the purpose of this paper is not to give a complete treatment of the anomaly issue in shape dynamics. This is a complicated problem which is still open in, for example, Loop quantum gravity. We would like to present what we feel is a possible setting and set of techniques to start an investigation of this matter. At the end of the conclusions we include a list of open issues we have left.

We start in the next section by giving a brief review of shape dynamics, its equations of motion in the asymptotically flat, maximal slicing case and its BRST charge. We then also briefly review some aspects of the Wess-Zumino consistency conditions, which gives a complete geometric classification of Weyl anomalies, begun by Bonora et al in \cite{Bonora-Cotta-Ramusino, Bonora-Pasti-Bregola, Bonora-Pasti-Tonin} and completed by Boulanger in \cite{Boulanger_general_WZ, Boulanger_algebraic_Weyl}. Following this introduction to the setting in which we will be working, we give a short introduction to Barnich's Hamiltonian formulation of anomalies. After making a number of assumptions about the translation of some basic results of Bonora et al and Boulanger, we can begin a study regarding which anomalous terms can arise in shape dynamics. In the last section, we apply these methods  and arrive at the results mentioned in the abstract. 

\subsection{Shape Dynamics}\label{sec:SD}

Here we will give a very brief construction of shape dynamics. We will cheat a little bit, and ignore the distinction in the construction of the theory between the spatial closed manifold case and the open manifold case. In one sense the open manifold case is simpler \cite{Asymptotic_SD} because it does not require the restriction of conformal transformations to be \emph{total volume preserving}, as it occurs in the closed manifold case \cite{SD_first}. On the other, the open manifold case is indeed  slightly more complicated because of the addition of boundary terms. Here we will use full conformal transformations and  also ignore the  boundary terms. The subject of our study here is insensitive to these omissions.

The first step in the construction of Shape Dynamics is to  write out the constraints of canonical GR in its 3+1 ADM form: 
\begin{eqnarray}
\label{equ:scalar constraint}S(x):= \frac{G_{abcd}\pi^{ab}\pi^{cd}}{\sqrt g}(x)-R(x)\sqrt g(x)=0\\
\label{equ:momentum constraint} H_a:={\pi^{a}_b}_{;a}=0
\end{eqnarray}
where the points $x$ belong to an open 3-manifold $\Sigma$, $g_{ab}$ is the spatial 3-metric and its conjugate momenta $\pi^{ab}$ (intimately related to the extrinsic curvature of a foliation). The scalar constraint \eqref{equ:scalar constraint} generates on-shell refoliations of spacetime, while the momentum constraint generates foliation preserving diffeomorphisms. The second step is to perform a canonical transformation in an extended phase space with coordinates $(g_{ab},\pi^{ab},\phi, \pi_\phi)$. The canonical transformation is of the form: 
$$(g_{ab},\pi^{ab},\phi,\pi_\phi)\mapsto (e^{4\phi}g_{ab},e^{-4\phi}\pi^{ab},\phi,\pi_\phi-4\pi)$$
where $\pi=g_{ab}\pi^{ab}$. We have an extra first class constraint in this extended theory, which is generated by $\pi_\phi-4\pi\approx 0$. This constraint generates a symmetry of the transformed variables.

The scalar constraint \eqref{equ:scalar constraint} becomes, for $\phi=\ln\Omega$  
\be\label{LY} \nabla^2\Omega+R\Omega-\frac{1}{8}\pi^{ab}\pi_{ab}\Omega^{-7}=0 
\ee
Ignoring boundary terms (see \cite{Asymptotic_SD} for details on how to treat the boundary terms), the smeared diffeomorphism constraint becomes
\be\label{equ:conf_diffeo_constraint}
H_a(\xi^a)=\int_\Sigma \left(\pi^{ab}\mathcal{L}_\xi g_{ab}+\pi_\phi\mathcal{L}_\xi \phi \right) d^3 x
\ee

Now one performs the gauge-fixing $\pi_\phi=0$ on this extended system. The only constraint that is second class with respect to this gauge-fixing is exactly \eqref{LY}. This constraint can be solved for $\Omega$ \cite{York} and the system reduced to a system with the canonical Poisson brackets of the variables $(g_{ab}\pi^{ab})$. 

It turns out that there remains a total Hamiltonian residing on the 
reduced phase space. After reduction we obtain the following set of constraints: 
obtain the constraints
\begin{equation}
 \begin{array}{rcl}
  D(\rho)&=&\int_\Sigma d^3x \,\rho \pi\\
  H_a(\xi^a)&=&\int_\Sigma d^3x \pi^{ab}(\mathcal L_\xi g)_{ab},
 \end{array}
\end{equation}
which generate unrestricted spatial diffeomorphisms and conformal transformations. The evolution in time is generated by the left-over physical Hamiltonian
\begin{equation}
 H=V-\int_\Sigma d^3x \sqrt{|g|}\Omega_o^6[g,\pi].
\end{equation}
where $\Omega_o$ is the solution to \eqref{LY}.
The non-zero part of the constraint algebra is given by:
\begin{eqnarray}
\{H_a(\eta^a),H_b(\xi^b)\}&=&H_a([\vec\xi,\vec\eta]^a)\nonumber\\
\label{equ:constraintAlgebra}
\{H^a(\xi_a),\pi(\rho))\}&=&\pi(\mathcal{L}_\xi\rho)
\end{eqnarray}
which substantially simplifies the algebra of constraints of gravity if compared to the ADM constraint algebra

\subsubsection{Equations of motion}
 Shape Dynamics is  obtained from a Linking theory \cite{SD:LT} (what we have called the extended theory), where all quantities are local. Thus the simplest way to  formulate Shape Dynamics' equations of motion, boundary charges, counter-terms and fall-off conditions is to consider these in the larger setting of the Linking theory, and then use phase space reduction.

 For the open case, where we institute maximal slicing as opposed to constant mean curvature slicing (CMC) which we institute when the spatial manifold is closed, the canonical transformation of the metric variables are given by $(g_{ab},\pi^{ab})\mapsto (e^{4\phi}g_{ab},e^{-4\phi}\pi^{ab})$. What makes the equations of motion so tractable in this case is that the conformal factor $\phi$ does not depend on the metric, as it does in the CMC case, where it is required to be total volume preserving. 

As in the CMC case, although the lapse does not figure in the fundamental equations, wherever it appears on the projection it is replaced by $N_o$, the solution of 
\be\label{LFE} e^{-4\phi_o}(\nabla^2 N_o+2g^{ab}\phi^o_{,a}N^o_{,b}) -N_oe^{-6\phi_o} G_{abcd}\pi^{ab}\pi^{cd}=0
\ee
where we have denoted the solution of \eqref{LY} by $\Omega_o[g,\pi]=e^{\phi_o[g,\pi]}$. 

The equations valid in the present case are:
\begin{eqnarray}
\label{equ:eom_g}\dot g_{ab}&=&4\rho g_{ab}+2e^{-6\phi_o}\frac{N_o}{\sqrt g}\pi^{ab}+\mathcal{L}_{\xi}g_{ab}\\
\dot\pi^{ab}&=& N_oe^{2\phi_o}\sqrt{g}\left(R^{ab}-2\phi_o^{;ab}+4\phi_o^{,a}\phi_o^{,b}-\frac{1}{2}R g^{ab}+2 \nabla^2\phi_o g^{ab}\right)\nonumber\\
&~&-\frac{N_o}{\sqrt g}e^{-6\phi_o}\left(2(\pi^{ac}\pi^b_c)-\frac{1}{2}(\pi^{cd}\pi_{cd})g^{ab}\right)\nonumber\\
&~&-e^{2\phi_o}\sqrt {g}\left(N_o^{;ab}-4\phi_o^{(,a}N_o^{,b)}-\nabla^2N_o g^{ab}\right)+\mathcal{L}_{\xi}\pi^{ab}-4\rho\pi^{ab}\label{equ:eom_pi}
\end{eqnarray}
Note the presence of the conformal gauge terms $4\rho g_{ab}$ and  $-4\rho\pi^{ab}$. One can explicitly check that $\{H, \pi\}=0$ from these equations of motion (and using the defining equations of $\Omega$ and $N_o$).

\subsubsection*{ The BRST variation for Shape Dynamics}

In the Hamiltonian formalism, now that we have gotten rid of the scalar ADM constraint, the symmetry transformations of our theory acts \emph{individually on each hypersurface}, which is a stronger condition than foliation preserving. The BRST Hamiltonian derivative $\delta$ for irreducible (first class\footnote{The BRST symmetries are only well-defined for first class systems.}) constraints, forming a rank one system  is given by $\delta= \xi^a\chi _a-\frac{1}{2}\xi^b\xi^aU_{ab}^c P_c$
where summation includes integration in the case of continuous variables,  $P^c$ are the ghost momenta associated to the first class constraints and $U_{ab}^c$ are the structure functions  for the first class constraints $\{\chi_ a\} $. In this condensed abstract index notation, subscripts stand in for  both the continuous and discrete variables. Note that the symmetries of the ghost fields are compatible with the  symmetries of the structure functions, so that $\eta_1\eta_2U_{12}^c =\eta_2\eta_1U_{21}^c$, for two constraints enumerated by $1$ and $2$. 

The only non-zero  elements of the Shape Dynamics constraint algebra matrix are:
\begin{eqnarray*} 
  U_{\pi(x), H_a(y)}^{\pi(z)}&=&\delta(z,y)_{;a}\delta(z,x)\\
 U_ {H_a(x), H_b(y)}^{H_c(z)}&=&\delta^c_b\delta(z,x)\delta(z,y)_{;a}-\delta^c_a\delta(y,z)\delta(z,x)_{;b}
\end{eqnarray*} 
Thus we can write for the BRST differential:
\begin{equation}\label{equ:BRST_SD}
 \delta=\int d^3x \left(\eta \pi+\xi^ag_{ac}\pi^{cd}_{;d}+\xi^b\xi^a_{,b}P_a+\frac{1}{2}\xi^a\eta_{,a}P\right)
\end{equation}
Here the ghosts associated with the Weyl constraints $\pi(x)=0$ are $\eta(x)$, while the ones associated with the momentum constraints are $\xi^a(x)$ (with respective ghost momenta $P$ and $P_a$). Note that here we are using $\xi^a$ as a ghost vector field, and not as the previously used bosonic vector field.  An explicit calculation shows that this definition for \eqref{equ:BRST_SD} yields $ \delta^2=0$. The last two terms in the differential only act on functions of non-zero ghost number. Thus for the first two terms in the chain of descent we can use the usual definition of the nilpotent differential generating Weyl transformations and diffeomorphisms. 

I.e. it is easy to see that we can split the action of \eqref{equ:BRST_SD} into:
\be\label{equ:Weyl_BRST_SD} s_{\mbox{\tiny W}}=\int\left( \eta g_{ab}\diby{}{g_{ab}}+  \eta \pi^{ab}\diby{}{\pi^{ab}}+\frac{1}{2}\eta_{,a}P
\diby{}{P_a}\right)
\ee
which matches the definition of the Weyl BRST variation with an extra field $\pi^{ab}$, \footnote{The last term comes from the fact that the conformal diffeomorphism ``group" is not a group, but a semi-group.} and 
\be\label{equ:Diffeo_BRST_SD} s_{\mbox{\tiny D}}= \int\left( \mathcal{L}_{\xi} g_{ab}\diby{}{g_{ab}}+   \mathcal{L}_{
\xi}\pi^{ab}\diby{}{\pi^{ab}}+\mathcal{L}_{
\xi}\eta\diby{}{\eta}+\mathcal{L}_{
\xi}\xi^a\diby{}{\xi^a}+\mathcal{L}_{
\xi}P\diby{}{P}+\mathcal{L}_{
\xi}P_a
\diby{}{P_a}\right)
\ee
where we have denoted $\xi$ as the index-free version of the ghost vector field $\xi^a$. To write out explicitly the new terms of \eqref{equ:Diffeo_BRST_SD}, remembering that $P$ and $P_a$ are ghost densities, we have: 
$$ \mathcal{L}_{\xi}\eta=\xi^a\eta_{,a}, ~~\mathcal{L}_{
\xi}\xi^a= \xi^b\eta_{,b}^a, ~~ \mathcal{L}_{
\xi}P= \frac{1}{2}(\xi^a_{,a}P+ \xi^a P_{,a}), ~~\mathcal{L}_{
\xi}P_a= \eta_ {,a}^bP_b+\xi^b_{,b}P_a+\xi^bP_{a,b}
$$

In the following sections of this paper we will try to adapt the Barnich Henneaux method of calculating the gauge cohomology in the Hamiltonian setting, and then resort to the cohomological manner of calculating Weyl anomalies introduced in \cite{Bonora-Cotta-Ramusino}.

\section{Cohomological treatment of the anomalies}\label{sec:Weyl}

The most adaptable formalism to find anomalies in our setting, is the geometric formulation of the consistency conditions. It should be stressed that this is a purely cohomological problem: i.e. we aim to identify only \emph{candidates} for the anomaly, a set which is dependent only on the field and symmetry content of the theory.\footnote{ The realm of applicability of this method is usually restrained to that of local field theories. In our case, although the symmetry generators are local, and we can compute fairly explicitly the equations of motion,  the Hamiltonian is not local. However, since the explicit form of the Lagrangian is not needed in the cohomological computation, we feel that our extension of the method to this form on non-locality is fairly reasonable. Furthermore, the restriction of locality pertains mostly to non-locality in time, and the non-invertibility of the operator $\square$. The case of spatial non-locality (and the respective invertibility of $\Delta$) should not present a major hurdle to the extension of the proofs.} The consistency conditions themselves (or the descent equations derived from them) are not capable of identifying the normalization of the anomalies, for which one must resort to other procedures, which do depend on the specific form of the action functional. In gauge theories these procedures can be: i) Fujikawa's covariant regularization in which the anomaly appears as a result of the transformation of the field Jacobian, ii) explicit perturbation theory, in which it is enough in 4 dimensions to calculate the triangle diagram (since the Adler-Bardeem theorem guarantees that there are no radiative corrections) iii) topological analysis, by using the Atyiah-Singer index theorem.

\subsection{Wess-Zumino consistency conditions}\label{sec:WZ}

To first illustrate the geometric character of the method, we will look at an abstract field theory with a symmetry generator.  Let us suppose that the field content is given by  $\Phi$, with arbitrary tensorial structure, on the manifold $M$.  We call the space of field configurations $\mbox{Sp}[\Phi]$, and the space of all gauge transformations $\mathcal{G}=\{g(x)\in G\}$. The physical space of the gauge theory is thus given by $\mbox{Sp}[\Phi]/\mathcal{G}$. 

The Wess-Zumino consistency conditions is nothing but the nilpotency condition of the vertical exterior derivative on $\mbox{Sp}[\Phi]$. Alternatively, the conditions can also be expressed as the commutation relations of variations along fundamental vector fields in $\mbox{Sp}[\Phi]$. We make this explicit in the following. 

 A variation of a functional $F[\Phi]$ is given by
$$\delta F[\Phi]=\int dx \delta \Phi(x)\diby{F}{\Phi(x)}
$$
which can be seen as a one form in $\mbox{Sp}[\Phi]$. Given a group action on $\Phi$, denoted by $g\cdot \Phi$, a fundamental vector field is given by $X[\Phi(x)]=\frac{d}{dt}(g_t\cdot \Phi(x))$.
For example, in the case of a gauge group $G$ with Lie algebra Lie $G$ and $\Phi=A_\mu^a$ we have $g_t\cdot\Phi(x)=A_\mu^a+t(D_\mu\xi)^a$ where $D$ is the gauge covariant derivative and $\xi^a\in $ Lie $G$. Given the basis $\diby{}{\Phi(x)}$ for $T\mbox{Sp}[\Phi]$ and $\delta \Phi(x)$ for $T^*\mbox{Sp}[\Phi]$, we can express the fundamental vector field associated to $\xi^a$ as 
$$ X_\xi= \int dx (D_\mu\xi)^a(x)\diby{}{A^a_\mu(x)}
$$
The Lie derivative of $F[A^a_\mu]$ along $X_\xi$ is then given by:
\be\label{equ:Lie_derivative} \mathcal{L}_{X_\xi}F[A^a_\mu]=\frac{d}{dt}(F[A_\mu^a+t(D_\mu\xi)^a])= \int dx (D_\mu\xi)^a(x)\diby{F}{A^a_\mu(x)}
\ee
The \emph{anomalous Ward identity} is defined as 
$$ \mathcal{L}_{X_\xi}W[A_\mu]=\imath _{X_\xi}\delta W[A_\mu]=:G(\xi, A)
$$
and the \emph{Wess-Zumino consistency condition}, a mere consequence of $\mathcal{L}_{X_\xi}\mathcal{L}_{X_\eta}-\mathcal{L}_{X_\eta}\mathcal{L}_{X_\xi}=\mathcal{L}_{X_{[\xi,\eta]}}$, which is also valid in this field space setting, can be written as:
\be\label{equ:WZ}
\mathcal{L}_{X_\xi}G(\eta, A)-\mathcal{L}_{X_\eta}G(\xi, A)=G([\xi,\eta], A)
\ee

We can write \eqref{equ:WZ} in more succinct form by defining \emph{the vertical exterior derivative in field space}. It can easily be seen that the exterior derivative of a gauge transformed $g\cdot A=:A_g$ decomposes as $\delta A_g=-D(g^{-1}\delta g)+g^{-1}\delta A g$, a vertical part and a part along the section $A$,\footnote{Chosen in the principal fiber bundle $\mbox{Sp}[A^a_\mu]$ over the base space $\mbox{Sp}[A^a_\mu]/\mathcal{G}$.}- which corresponds to the derivative along the moduli space $\mbox{Sp}[A^a_\mu]/\mathcal{G}$. Thus defining the vertical derivative as the derivative restricted to the fibers: 
$$\delta_{\mbox{\tiny{vert}}} A_g:={\delta A_g}|_{\mbox{\tiny{fiber}}}=-D(g^{-1}\delta g)=-D\omega
$$
where we recognize the \emph{Maurer-Cartan form} $\omega:=g^{-1}\delta g$. This formula automatically gives $\delta_{\mbox{\tiny{vert}}}\omega=-\omega^2$ and $\delta_{\mbox{\tiny{vert}}} d=d \delta_{\mbox{\tiny{vert}}}$. We can now define the one-form anomaly as $\delta_{\mbox{\tiny{vert}}} W$, which immediately yields the concise formula of the Wess-Zumino consistency condition: 
\be\label{equ:WZconcise}
\delta_{\mbox{\tiny{vert}}} G(A)=\delta_{\mbox{\tiny{vert}}}^2 W[A]=0
\ee
We can rewrite the ``one-forms" in group space as Fadeev-Popov ghosts: $\omega=v_a T^a$, where $T^a$ is a Lie algebra basis and $v^a$ are the anti-commuting FP ghosts.\footnote{Note that just as the usual, finite-dimensional Maurer-Cartan form $g^{-1}dg$ generates the Lie algebra $\mathfrak{g}$, in this case $\omega=g^{-1}\delta g$ generates the one-forms on the Lie algebra space.}  This yields the more common expressions for the BRST variation of the FP ghosts $\delta_{\mbox{\tiny{vert}}} v^a=-\frac{1}{2}[v,v]^a$. It also becomes easier to express the ``one-form" (i.e. ghost number 1) given by the vertical exterior derivative of a functional of the fields:
 \be\label{equ:vertical_derivative_FP}
 \delta_{\mbox{\tiny{vert}}} F[A]=\int dx( D_\mu v)^a\diby{F}{A^a_\mu(x)}
 \ee
which differs from \eqref{equ:Lie_derivative} merely by the substitution of the Lie algebra element $X^a$ by the FP ghost (one form) $v^a$. Augmented by the variation $\delta_{\mbox{\tiny{vert}}} v^a=\frac{1}{2}[v,v]^a$ we generate the full structure of a graded Lie algebra (in ghost number).  

It can be shown \cite{Bertlmann} that $G(A)=\delta_{\mbox{\tiny{vert}}} W[A]$ is a local functional of the fields, i.e. is the integral of a given function of the fields.\footnote{A recipe exists for deriving non-local extensions of the anomaly, both in the consistent version - used here - or in the covariant version. But to derive such an extension one must start with the local version. }  A true anomaly however will correspond to a $G(A)$ that cannot be removed by the regularization procedure. That is, it cannot be removed by the addition of a local counter-term to the effective action:
$$ W[A]-\int dx f(A)(x)=W[A]-f(A)
$$
where we denote the integral by dropping dependence on $x$. Thus a true anomaly will correspond to a term $\delta_{\mbox{\tiny{vert}}} h(A)= 0$ but such that $h(A)\neq \delta_{\mbox{\tiny{vert}}} f(A)$, that is, \emph{the anomaly is given purely by a cohomology problem}.



To adapt the problem to the 3+1 setting, we must first introduce the \emph{Stora-Zumino chain of descent}, which is an iteration of the Wess-Zumino consistency conditions coupled to both successive exterior differentiation and use of the Poincar\'e lemma to obtain recursion relations  between the possible anomaly terms.  
An easy procedure to obtain the Stora-Zumino chain of descent is the following: let locally 
 \be
 \label{local_WZ}\delta_{\mbox{\tiny{vert}}} h(A)= \int dx \delta^{\mbox{\tiny{loc}}}_{\mbox{\tiny{vert}}} h(A)(x)=0
 \ee
where $h(A)(x)$ is ghost number one $n$-form, let us call it $Q^n_1$. Equation \eqref{local_WZ} implies  
\be \label{equ:local_WZ}\delta^{\mbox{\tiny{loc}}}_{\mbox{\tiny{vert}}}Q^n_1 (x)+d Q_2^{n-1}(x)=0\ee where $Q_2^{n-1}(x)$ is an $n-1$ form with ghost number $2$. By applying to this expression $\delta^{\mbox{\tiny{loc}}}_{\mbox{\tiny{vert}}}$ again, using nilpotency,  $\delta^{\mbox{\tiny{loc}}}_{\mbox{\tiny{vert}}} d=d\delta^{\mbox{\tiny{loc}}}_{\mbox{\tiny{vert}}}$ and the Poincar\'e lemma, we obtain $$\delta^{\mbox{\tiny{loc}}}_{\mbox{\tiny{vert}}} Q_2^{n-1}(x)= d  Q_3^{n-2}(x)$$ and so on, until $\delta^{\mbox{\tiny{loc}}}_{\mbox{\tiny{vert}}}  Q_n^{0}(x)=0$, the bottom of the  well-known \emph{Stora-Zumino chain of descent equations}, which is the main tool used in finding anomalies (up to normalization) in arbitrary dimensions. The strategy is  to first find the bottom of the descent equation and then look for obstructions to its lifting.\footnote{A common strategy is to look first for a maximal degree form in $n+2$ dimensions that is both closed and gauge-invariant. In gauge theories this is usually taken to be a Chern polynomial $P(F^{n+1})$. Since it is closed, $P(F^{n+1})=dQ^0_{2k+1}$, since it is invariant $\delta^{\mbox{\tiny{loc}}}_{\mbox{\tiny{vert}}}Q^0_{2k+1}=dQ^1_{2k}$, where $Q^1_{2k}$ is the anomaly (which clearly satisfies $\int\delta^{\mbox{\tiny{loc}}}_{\mbox{\tiny{vert}}}Q^1_{2k}=\int dQ^2_{2k-1}=0$).} From now on we will denote $\delta^{\mbox{\tiny{loc}}}_{\mbox{\tiny{vert}}}$ by the operator $s$. In this notation one can characterize the descent equations as providing a homomorphism $\mathcal{D}:H^{(g,k)}(s|d)\rightarrow H^{(g+1,k-1)}(s|d)$ with $\mathcal{D}[a]=[b]$ where $a$ and $b$ satisfy $sa+db=0$ and $sb+dc=0$, for some lower $c$, and so on. 

\subsubsection{Wess-Zumino consistency for the Weyl anomaly}

In a theory that is classically diffeomorphism and Weyl invariant, a BRST treatment will give rise to a BRST differential $s=s_{\mbox{\tiny W}}+s_{\mbox{\tiny D}}$, where $s_{\mbox{\tiny D}}$ generates the BRST transformation associated to the diffeomophisms, and $s_{\mbox{\tiny W}}$ generates those  associated to Weyl transformations. If we are dealing with the purely gravitational problem in space-time, apart from the space-time metric $g_{\mu\nu}$, the only other fields in consideration are the Weyl ghost scalar $\eta$ and the ghost vector field $\xi^\mu$, both anti-commuting variables with ghost number 1. In our case, we will have to add the momentum variables to this set. 

The non-trivial action of the $s$ on the set of variables above is given by: 
\begin{eqnarray}
\label{1}s_{\mbox{\tiny D}}g_{\mu\nu}= \mathcal{L}_\xi g_{\mu\nu}\\
s_{\mbox{\tiny D}}\eta= \mathcal{L}_\xi\eta=\xi^\rho\eta_{,\rho}\\
s_{\mbox{\tiny D}}\xi^\mu=\mathcal{L}_\xi\xi^\mu=2\xi^\rho\xi_{,\rho}^{\mu}\\
\label{2}s_{\mbox{\tiny W}}g_{\mu\nu}=2\eta g_{\mu\nu}
\end{eqnarray}
the remaining actions being zero. Note that here we are adopting the scaling $2$ instead of $4$ for the metric, which is the usual scaling for a space-time action in 4-dimensions. Equation \eqref{equ:local_WZ} for ghost number 1 implies in this case, for $s=s_{\mbox{\tiny D}}+s_{\mbox{\tiny W}}$, that: 
\be\label{equ:WZ_coboundary} s a^n_1+db^{n-1}_2=0, ~~~~ a^n_1\neq s p^n_0+dq^{n-1}_1
\ee
where the superscript denotes the form degree and subscript denotes the ghost number. 

For Weyl anomalies, we look for elements whose ghost number 1 comes from the Weyl ghost. Decomposing \eqref{equ:WZ_coboundary} according to Weyl ghost number we obtain the equations: 
\begin{eqnarray}
s_{\mbox{\tiny D}}a_1^n+db_2^{n-1}=0\label{equ:decomposed_WZ1}\\
s_{\mbox{\tiny W}}a_1^n+dc_2^{n-1}=0,~~a^n_1\neq  s_{\mbox{\tiny W}}p^n_0+dq^{n-1}_1,~~s_{\mbox{\tiny D}}p^n_0+df^{n-1}_1=0\label{equ:decomposed_WZ2}
\end{eqnarray}
where the last equation of \eqref{equ:decomposed_WZ2} means that we can restrict the coboundary conditions to terms that are diffeomorphism invariant. In other words, the last equation of \eqref{equ:decomposed_WZ2} and \eqref{equ:decomposed_WZ1} allow us to  compute the cohomology
$H^{(1,n)}(s_{\mbox{\tiny W}}, | d)$ of the Weyl BRST differential $s_{\mbox{\tiny W}}$ modulo total
derivatives, restricted to the space of diffeomorphism-invariant
local n-forms.  It is shown in \cite{Bonora-Pasti-Bregola} that  by adding a local
 counterterm to the action,  one can always shift 
the pure diffeomorphism part of the candidate anomaly,
 leaving only the pure Weyl $a^n_1$. We will use this in the restriction of the possible terms in the anomaly search of section \ref{sec:Weyl}. 
 
 Boulanger is able to write a general classification theorem for Weyl anomalies using the above formalism and a ``Weyl-covariant" calculus. For us, the important results are that the cohomology vanishes for odd dimensions, and that  $H^{(2,3)}(s_{\mbox{\tiny W}}, | d)$ has a single generator, given by:
 \be\label{equ:Schwinger}\tilde b^{(2,3)}_0= -2\sqrt{g}\eta \epsilon^{\alpha}_{\nu\rho\sigma}\eta_{,\alpha}(R^\nu_\mu -\frac{1}{4}\delta^\nu_\mu )dx^\mu dx^\rho dx^\sigma
 \ee

\subsection{Hamiltonian treatment of the WZ cross-consistency conditions}\label{sec:Hamiltonian_WZ}

In order to apply the formalism to a Hamiltonian setting, we adapt the results obtained by Barnich in \cite{Barnich_Hamiltonian} and Barnich and Henneaux in \cite{Barnich&Henneaux}. In the absence of these results, the issues that could be present  in attempting to translate directly the WZ consistency conditions are related to the fact  that the Hamiltonian setting uses as main variables the momenta and positions variables, but the action - through which the usual WZ conditions are expressed - includes time derivatives and Lagrange multipliers.

The manner in which Barnich and Henneaux translate the Lagrangian BRST cohomology into the Hamiltonian one,  is by first building a map from the Batallin-Vilkovisky (BV) anti-bracket to the Poisson bracket. In the BV formalism for irreducible gauge theories \cite{Gomis_antifield}, besides the original fields of ghost number zero the theory $\phi^\mu$, and the the ghosts $\eta^\alpha$ of ghost number one, one introduces anti-fields $\phi^*_\mu$ and $\eta^*_\alpha$ of ghost numbers $-1$ and $-2$ respectively. Generalizing notation to $\phi^A=(\phi^\mu, \eta^\alpha)$ one then declares the variables $\phi^A$ and $\phi_A^*$ as conjugate in an extended \emph{anti-bracket}, which can then be trivially extended to arbitrary functions of $\phi^A$ and $\phi_A^*$. By contrast, in the Hamiltonian framework one has the conjugate spatial position and momentum fields $q^a$ and $p_a$ resp., and the conjugate spatial ghosts and ghosts momenta $\xi^a$ and $P_a$, and an extended Poisson bracket by which they are canonical (as used above in order to find the Hamiltonian BRST charge for shape dynamics \eqref{equ:BRST_SD}).  The ghost momenta are intimately related to the anti-fields (they can in fact be identified with the anti-field of the Lagrange multiplier associated to the first class constraints). 

The isomorphism is highly non-trivial, and we will thus avoid giving a complete description of it. The first fact of importance, as derived in \cite{Barnich&Henneaux}, is that the local BRST cohomology group is invariant with respect to the introduction of generalized auxiliary fields (such as Lagrange multipliers). One can then construct, out of the irreducible first class constraints in the Hamiltonian formalism, the BRST charge (e.g. \eqref{equ:BRST_SD}), which we will call for generality 
$$\Omega=\int d^{n-1}x\omega$$
which is a local functional \emph{in space} satisfying $\{\Omega, \Omega\}=0$. For the work of Barnich and Henneaux, the Hamiltonian is also assumed to  be a local functional in space, of the form 
$$H=\int d^{n-1}x h.$$
Both of these functionals depend solely on the fields and their conjugate momenta and their spatial derivatives (no time derivatives or anti-fields in these objects). 

In our case the Hamiltonian is not local. Its symplectic flow around a flat solution can be put in a  local form\footnote{To wit, the local form around a flat solution, using the flat Laplacian Green's function is given by:  $$\chi_H(x)=g^{ab}(x)\diby{}{g_{ab}(x)}+\int d^3 x' \frac{\nabla_a\nabla_b\diby{}{g_{ab}(x')}}{|r-r'|}.$$ }and there are also local derivative expansions for the Hamiltonian \cite{Tim_effective}, but  it is not local nonetheless. Having said this, as has been pointed out to us by Boulanger in private communication, the sort of strictly spatial non-locality (through the invertible spatial Laplacian $\Delta$) that shape dynamics enjoys is unlikely to do any harm to the formal proofs of the statements used here, it is non-locality in time that is the more harmful one. 

In any case, the first important result for us is theorem 6.1 in \cite{Barnich&Henneaux},
\begin{theo}[Barnich and Henneaux, 6.1]\label{Theo:1}
The ordinary BRST cohomology depending on the fields $\phi^A$, the antifields $\phi^*_A$ and their space-time derivatives is isomorphic to the cohomology of $s_\omega$, depending only on the spatial Poisson-bracket-conjugate fields $\phi^i, \pi_i $ and $\eta^i$ and $P_i$ and their spatial derivatives.
\end{theo}
 which shows that the ordinary BRST cohomology depending on the full spacetime fields and anti-fields is isomorphic to the cohomology of $s_\omega$ -- the operator associated to the BRST charge $\omega$, $s_\omega=\{\omega, \cdot\}$ -- depending only on the \emph{spatial fields} of the Hamiltonian formalism (and their spatial derivatives).  The complete BRST symmetry of the Hamiltonian system (which is isomorphic to the Lagrangian BRST symmetry after elimination of the auxiliary fields) includes more than just the BRST charge associated to the first class constraints, it is generated by a full solution of the master equation $(S_H, S_H)=0$ in the anti-bracket formalism. The importance of theorem \ref{Theo:1} is that is shows that one can get rid of the temporal derivatives and of the anti-fields through the addition of an $s_H= \{S_H, \cdot\}$ coboundary. 

Then, by starting at the bottom of the descent equations, as in \ref{sec:WZ}, using successive shuffling of derivatives -- which one can do since the cohomology is modulo the full spacetime exterior derivative $d$ -- and separating out the $dt$ part of a given $n$-form, they show that  a non-trivial cocycle  modulo $d$ is algebraically given by: 
\be a^{(g,k)}=dt(\{\phi^A\phi^*_A,\tilde{b}^{(g+1,k-1)}_0\}+a^{0(g,k-1)}_0)+\tilde{a}^{(g,k)}_0
\ee
 and the cocycle components used above satisfy: 
\begin{eqnarray}
\label{equ:new_cocycle1}s_\omega\tilde{a}^{(g,k)}_0+\tilde{d}\tilde{b}^{(g+1,k-1)}_0=0\\
\label{equ:new_cocycle2}s_\omega a^{0(g,k-1)}_0+\tilde{d}b^{0(g+1,k-2)}_0-\frac{\partial}{\partial t}\tilde{b}^{(g+1,k-1)}_0+\{H,\tilde{b}^{(g+1,k-1)}_0\}=0
\end{eqnarray}
where we use tildes to denote purely spatial quantities, i.e. $\tilde{a}_0 , a^0_0, \tilde{b}_0 $ and $b^0_0$ contain no antifields and no time derivatives of the fields, and the superscript parentheses $(g,k)$ denotes the ghost number ($g$) and differential form degree $(k)$ respectively.\footnote{One difference in this framework to what is done in \cite{Barnich&Henneaux} is that they insist on a local bracket, i.e. between two local quantities. For the formal definition to satisfy the usual requirements of a bracket (it does only up to boundary terms), they define a bracket by using local Euler-Lagrange derivatives. For us we will not need this because i) we are happy to use the bracket with the integral of $h$, and not $h$ itself and ii) all the quantities considered here do not depend on more than second derivative of the metric and zero derivatives of the momenta, in which case the use of the Euler-Lagrange derivation is superfluous.}  As in the usual descent equations, $\tilde{b}_0 $ and $b^0_0$ satisfy analogous equations  to \eqref{equ:new_cocycle1}, \eqref{equ:new_cocycle2}, with respect to some new $\tilde{m}_0 $ and $m^0_0$, and so on. In maximum space-time form degree there is of course no $\tilde{a}$, and at the bottom of the descent equations, let's say  at $k$ steps, $\tilde{k}_0 $ and $k^0_0$ are $s_\omega$ - cocycles. 

Under these modified descent equations, we have slightly different coboundary conditions (from \eqref{equ:new_cocycle1}, \eqref{equ:new_cocycle2}). For $a^{(g,k)}= s_Hc^{(g-1,k)}+d e^{(g,k-1)}$, a decomposition of $c$, 
$$ c^{(g-1,k)}=dt(\{\phi^A\phi^*_A,\tilde{e}^{(g,k-1)}_0\}+c^{0(g-1,k-1)}_0)+\tilde{c}^{(g-1,k)}_0
$$
yields:  
\begin{eqnarray}
\label{equ:coboundary_new1}\tilde a^{(g,k)}_ 0=s_\omega\tilde{c}^{(g-1,k)}_0+\tilde{d}\tilde e_0\\
\label{equ:coboundary_new2} a^{0(g,k-1)}_0=-s_\omega c^{0(g-1,k-1)}_0-\tilde{d}e^{0(g,k-2)}_0+\frac{\partial}{\partial t}\tilde{e}^{(g,k-1)}_0+\{H,\tilde{e}^{(g,k-1)}_0\}
\end{eqnarray}

The important classification theorem derived from these equations is: 
\begin{theo}[Barnich and Henneaux, 6.2]\label{theo:2}
The local BRST cohomology groups are isomorphic to the direct sum of the following three local cohomology groups of the Hamiltonian formalism (for ghost number $g$ and in form dimension $k$):
\be \label{equ:decomposition}H^{(g,k)}(s_H | d)\simeq H^{(g,k)}(s_\omega | \tilde d)\oplus l[H^{(g+1,k-1)}(s_\omega | \tilde d)]\oplus r[H^{(g,k-1)}(s_\omega | \tilde d)]\ee
\end{theo} 
Let us explain the notation. To characterize the full cohomology $H^{(g,k)}(s_H | d)$, one finds first a time-independent, antifield independent basis for the space $H^{(g,k)}(s_\omega | \tilde d)$, which represents the most general solution for $\tilde a_0$. On maximal form dimension this is zero, since it cannot depend on $dt$. Then one does the same for $H^{(g+1,k-1)}(s_\omega | \tilde d)$, finding a basis represented by  $\tilde b^{(g+1,k-1)}_0$. The subspace $l[H^{(g+1,k-1)}(s_\omega | \tilde d)]$ is that for which equation \eqref{equ:new_cocycle2} admits a particular solution $a_{0P}^0$. The space $r[H^{(g,k-1)}(s_\omega | \tilde d)]$ characterizes the subspace of $H^{(g,k-1)}(s_\omega | \tilde d)$ which  obeys the new decomposed boundary \eqref{equ:coboundary_new2}. Let us explain a bit further these last two elements. They are obtained by a decomposition of $a^0_0$ into a ``homogeneous" solution  $\bar a_{0}^0$ , and a ``non-homogeneous" one $a_{0P}^0$. The non-homogeneous part is taken to any \emph{particular} solution of \eqref{equ:new_cocycle2}, and it is called non-homogeneous because, apart from the usual cocycle elements, it also contains $\frac{\partial}{\partial t}\tilde{b}^{(g+1,k-1)}_0+\{H,\tilde{b}^{(g+1,k-1)}_0\}$. The homogeneous part $\bar a_{0}^0$ then is taken to satisfy \eqref{equ:new_cocycle2} \emph{without} the presence of this extra term, but must also respect the coboundary equation  \eqref{equ:coboundary_new2}. Thus is just an element of $H^{(g,k-1)}(s_\omega | \tilde d)$ which also respects the new coboundary , i.e. it is an element of $r[H^{(g,k-1)}(s_\omega | \tilde d)]$. It is easy for example to see that if there is no  $\tilde{b}^{(g+1,k-1)}$ for which 
$a_{0P}^0$ exists (or if $\tilde{b}^{(g+1,k-1)}$ is zero) then the entire anomaly will come from the homogeneous part $\bar a_{0}^0$, which vanishes if $H^{(g,k-1)}(s_\omega | \tilde d)=0$ for example. 


Using these decompositions, Barnich (\cite{Barnich_Hamiltonian}, equations 26-27) finds  that the classical relations 
$$\{\Omega, \Omega\}=0 ~, ~~~\{ \Omega, H\}=0
$$
acquire anomalous terms:
\begin{eqnarray}
\int d^{n}x[\Omega, \Omega]=\frac{\hbar}{2}\int \tilde b^{(2,3)}_{0}  dt  +\mathcal{O}(\hbar^2)\label{equ:spatial_anomaly}\\
\int d^{n}x[\Omega, H]=\frac{\hbar}{2}\int a^{0(1,3)}_0 dt  +\mathcal{O}(\hbar^2)\label{equ:time_anomaly}
\end{eqnarray}
where we used square brackets for the quantization of the commutation relations. 
Now we move on and attempt to apply this to the shape dynamics scenario.

\subsection{The treatment for Shape Dynamics}

\subsubsection*{The calculation}

First let us rewrite the equations \eqref{1}-\eqref{2} for the BRST differentials \eqref{equ:Weyl_BRST_SD}  and  \eqref{equ:Diffeo_BRST_SD}. The non-zero transformations are:
\begin{eqnarray}
s_{\mbox{\tiny D}}g_{ab}= \mathcal{L}_\xi g_{ab},~&~&~
s_{\mbox{\tiny D}}\pi^{ab}= \mathcal{L}_\xi \pi^{ab}\\
s_{\mbox{\tiny D}}\eta= \mathcal{L}_\xi\eta, ~&~&~s_{\mbox{\tiny D}}P= \mathcal{L}_\xi P\\
s_{\mbox{\tiny D}}\xi^a=\mathcal{L}_\xi\xi^a, ~&~&~s_{\mbox{\tiny D}}P_a=\mathcal{L}_\xi P_a \\ 
s_{\mbox{\tiny W}}g_{ab}=\eta g_{ab},~&~&~s_{\mbox{\tiny W}}\pi^{ab}=-\eta \pi^{ab}
\end{eqnarray}
and the extra equation from the semi-group quality of the conformal diffeomorphisms: $s_{\mbox{\tiny W}}P_a=\frac{1}{2}\eta_{,a}P$.

The first assumption we will make is that although the shape dynamics Hamiltonian is not local, we can still apply the cohomological method to calculate certain aspects of the anomaly. This is reinforced  by the very limited dependence of the calculation on the explicit form of the action functional. We can apply this locality assumption at different levels. We will apply if first to find the appropriate cohomology spaces,\footnote{Without this assumption of course it is meaningless to work with the local cohomology spaces, and our search would be unfeasible.} but as we will see, this can still allow for some non-locality of the anomaly coming from the action of the Hamiltonian. At the end of the calculation, we are still able to impose the further condition of locality of the anomaly itself. 

The second assumption, is that for the calculation of a Weyl anomaly, a Bardeem-Zumino term can shift the diffeomorphism part of the anomaly to the Weyl part. That this goes through in the usual Lagrangian setting has been shown by Bonora et al \cite{Bonora-Pasti-Bregola}. 
 This will allow us to look for Weyl anomalies of \eqref{equ:Weyl_BRST_SD} among the spatially diffeomorphism invariant cocycles of \eqref{equ:Diffeo_BRST_SD}. Identifying the general $s_\omega$ from the previous section \ref{sec:Hamiltonian_WZ} with $s$ given implicitly in \eqref{equ:BRST_SD}, this assumption (which is equivalent in our setting to \eqref{equ:decomposed_WZ1}, \eqref{equ:decomposed_WZ2}) allows us to use $s_{\mbox{\tiny W}}$ in \eqref{equ:decomposition},
\be \label{equ:new_decomposition} H^{(g,k)}(s_{\mbox{\tiny W}} | \tilde d)\oplus l[H^{(g+1,k-1)}(s_{\mbox{\tiny W}} | \tilde d)]\oplus r[H^{(g,k-1)}(s_{\mbox{\tiny W}} | \tilde d)]
\ee
 for the  restricted  space of diffeomorphism-invariant
local n-forms. We should stress that although this a reasonable assumption -- specially so for the spatial part of the anomaly, where we do not expect any obstructions --  we have not tried to explicitly prove it.

The first thing we show is that the Weyl cohomologies for Shape Dynamics, as given in \eqref{equ:decomposition}, cannot depend  on the momenta, and by the results mentioned in the previous section, also cannot depend on the Lagrange multipliers or the time derivatives of the fields. 

  To start the calculation, we note that if the 3-metric $g_{ab}$ has conformal weight $4$, the symmetric 2-tensor $p_{ab}=g_{ac}g_{bd}\pi^{cd}/\sqrt g$ has conformal weight $-2$, and the totally anti-symmetric 3-tensor $\epsilon^{abc}$ has conformal weight $-6$. For Weyl ghost number one with no ghost momenta, namely for anomalies, we can shuffle derivatives and always leave the ghosts undifferentiated. Thus the assumption here is to take  our space of cochains as being formed by integrated local polynomials of the fields $g_{ab}, g^{ab},p^{ab}, \epsilon^{abc} $,  and  derivatives $\partial_a$. The polynomials  will be smeared by $\eta$ to form the actual ghost number one Weyl anomalies. 

Thus, assuming that the Bardeem-Zumino shifting of the anomalies works also in this context and thus we can restrict our search to the space of diffeomorphism invariant functionals, we must match the following tensor  indices and conformal weights:
\begin{eqnarray}
-3N_\epsilon+ 2N_p+2N_g-2N_{g^{-1}}+N_{\nabla}&=&0\label{equ:counting_tensor}\\
-6N_\epsilon - 2N_p+4N_g-4N_{g^{-1}}&=&-6\label{equ:counting_conf}
\end{eqnarray}
Multiplying the first line by 2 and subtracting the second line we obtain $6N_p+2N_\nabla=6$. So either $N_p=1$ and $N_\nabla=0$ or $N_p=0$ and $N_\nabla=3$. Let us pause to note here that consistently, if we were in $d$ spatial dimensions, we would obtain $N_p+d^{-1}N_\nabla=d^{-1}$ which matches the actual number of derivatives of the known geometrical anomalies. The important part of this calculation is that we can see that the momenta appear only when there is no derivatives of the metric involved. It can thus appear only as the term $\pi$, which vanishes on-shell and is thus not a candidate for the anomaly. This means that at least for Weyl ghost number one we can take the usual geometric notion of Weyl anomalies, which are classified up to 8 dimensions \cite{Boulanger_general_WZ, Boulanger_algebraic_Weyl}. For odd dimensions, $k=4$ in \eqref{equ:new_decomposition}, this  takes care of $r[H^{(1,3)}(s_{\mbox{\tiny W}} | \tilde d)]=0$, and since we are in maximal form dimension, we also have $H^{(1,4)}(s_{\mbox{\tiny W}} | \tilde d)=0$. This means that the homogeneous part of the time-anomaly, as given in \eqref{equ:time_anomaly}, vanishes in odd dimensions.

 The problem  is that the total anomaly as described by the decomposition \eqref{equ:decomposition}, involves also the ghost number 2 cohomology element in $l[H^{(2,3)}(s_{\mbox{\tiny W}} | \tilde d)]$. 
 To calculate the possible cohomology  terms it is not enough anymore to consider only the space of cochains as being formed by integrated local polynomials of the fields $g_{ab}, g^{ab},p^{ab}, \epsilon^{abc} $ and   derivatives $\partial_a$ because integration by parts cannot isolate the ghost terms. This definitely deserves further investigation, but the purpose of this paper is only to establish the necessary tools, and provide the appropriate setting to deal with a shape dynamics anomaly. To classify all the possible polynomial Schwinger terms goes beyond our scope. 

We want to illustrate the hurdles in performing the full calculation, so let us move forward by \emph{assuming} that even for $H^{(2,3)}(s_{\mbox{\tiny W}} | \tilde d)$ only the geometric (momentum-independent) terms arise. 

Following the prescription of theorem \ref{theo:2}, we can start with the second cohomology term \eqref{equ:Schwinger}:
\be\label{b32} \tilde b^{(2,3)}_0= -2\sqrt{g}\eta \epsilon^{\alpha}_{\nu\rho\sigma}\eta_{,\alpha}(R^\nu_\mu -\frac{1}{4}\delta^\nu_\mu )dx^\mu dx^\rho dx^\sigma
 \ee 
Note that this term has a four-index $\epsilon$-tensor, basically because it has descended from the 4-dimensional anomaly. Since we are in maximal form dimension, we have to find a non-trivial solution for $a^0_0$ to the non-homogeneous \eqref{equ:new_cocycle2}:
\be s_{\mbox{\tiny W}} a^{0(1,3)}_0+\tilde{d}b^{0(2,2)}_0-\frac{\partial}{\partial t}\tilde b^{(2,3)}_0+\{H,\tilde b^{(2,3)}_0\}=0\ee
and thus we must follow the chain of descent, so as to determine first $b^0_0$. That is, we have: 
\begin{eqnarray}s_{\mbox{\tiny W}} \tilde b^{(2,3)}_0 +\tilde{d}\tilde{m}^{(3,2)}_0=0\\
\label{b0}s_{\mbox{\tiny W}} b^{0(2,2)}_0+\tilde{d}m^{0(3,1)}_0-\frac{\partial}{\partial t}\tilde{m}^{(3,2)}_0+\{H,\tilde{m}^{(3,2)}_0\}=0
\end{eqnarray} 
and so on. Thus we start by calculating $s_{\mbox{\tiny W}}\tilde b^{(2,3)}_0$, in order to find $\tilde{m}_0$. We also note that there is no explicit time dependence in any of our spaces, since our $s_\omega=s_{\mbox{\tiny{D}}}+s_{\mbox{\tiny{W}}}$ is completely hypersurface intrinsic. Thus we set the explicit time-derivatives to zero. 

After some algebra we have that the Weyl variation of $\tilde b^{(2,3)}_0$ yields: 
\be s_{\mbox{\tiny W}} \tilde b^{(2,3)}_0= 2\eta\sqrt{g} \epsilon^\alpha_{~\nu\rho\sigma}  \partial_\alpha \eta\nabla_\mu \partial^\nu \eta dx^\mu dx^\rho dx^\sigma 
\ee
which after some manipulation can be written as a total derivative of a term $\tilde{m}^{(3,2)}_0$: 
\be s_{\mbox{\tiny W}} \tilde b^{(2,3)}_0= \frac{1}{2}\nabla_a\left(\eta\sqrt g \epsilon^{bc}_{~~de}\eta_{,b} \eta_{,c}dx^a dx^d dx^e\right)=\tilde d\tilde{m}^{(3,2)}_0
\ee
thus 
\be  \tilde{m}^{(3,2)}_0= -\frac{1}{2}\eta\sqrt g \epsilon^{bc}_{~~de}\eta_{,b} \eta_{,c} dx^d dx^e
\ee
With a little more algebra one can now check that this is indeed where the descent stops, i.e. $s_{\mbox{\tiny W}} \tilde{m}^{(3,2)}_0=0$. Thus  we have from the respective descent \eqref{equ:new_cocycle2} for the $m_0$ level that $\tilde{d}m^0_0=0$. Inputting this back in \eqref{b0} we are left with 
\be\label{3}s_{\mbox{\tiny W}} b^{0(2,2)}_0+\{H,\tilde{m}^{(3,2)}_0\}=0
\ee
However, from the equations of motion of shape dynamics \eqref{equ:eom_g}, since $\tilde{m}^{(3,2)}_0$ contains no derivatives of the metric, and depends on it only through $\sqrt g$,  we have that 
$$\{H,\tilde{m}^{(3,2)}_0\}\propto g^{ab}\dot{g}_{ab}\approx0$$
 and \eqref{3} becomes the bottom of the descent equations for the cohomology  $H^{(2,2)}(s_{\mbox{\tiny W}} | \tilde d)$ which can be checked to vanish \cite{Boulanger_general_WZ}. Thus  $b^0_0=0$, and we have all the necessary elements to characterize the shape dynamics Weyl anomaly in 3+1 dimensions. It will be characterized by a non-trivial solution of 
\be\label{final} s_{\mbox{\tiny W}} a^{0(1,3)}_0+ \alpha\{H,\tilde b^{(2,3)}_0\}=0\ee
where the action of the shape dynamics Hamiltonian $H$ is given by \eqref{equ:eom_g} and \eqref{equ:eom_pi}. If there is such a non-trivial solution for $a^0_0$ for $\alpha\neq 0$ there will be  candidate anomalies, both temporal and spatial.

We will refrain from writing our equation \eqref{final} in all of its gory detail. Although the only unknown in equation \eqref{final} is $a^0_0$, the equation itself is a complicated  functional differential equation with coefficients given in terms of geometric quantities (such as the Ricci curvature and the York scalar) and it is not our aim in this paper to perform a complete investigation on this matter. Our aim is to begin an exploration in  the methods and computations necessary to study the Weyl anomaly in the context of shape dynamics.

Thus under the present assumptions we have obtained that any anomaly can appear only as multiples of
\begin{eqnarray}
\int d^{n}x[\Omega, \Omega]=\frac{\alpha\hbar}{2}\int \tilde b^{(2,3)}_{0}  dt  +\mathcal{O}(\hbar^2)\label{equ:SD_spatial_anomaly}\\
\int d^{n}x[\Omega, H]=\frac{\alpha\hbar}{2}\int a^{0(1,3)}_0 dt  +\mathcal{O}(\hbar^2)\label{equ:SD_time_anomaly}
\end{eqnarray}
if there exists a solution $a^0_0$  for \eqref{final} with $\alpha\neq 0$, where $H$ is the shape dynamics Hamiltonian  and $\tilde b^{(2,3)}_0$ is given by \eqref{b32}. We should note however that so far we have only imposed the cohomology spaces to be formed by local terms. If we impose the further restriction that $a^{0(1,3)}_0$  should  be a local term, then there can be no solution, as a direct calculation shows that $\{H,\tilde b^{(2,3)}_0\}$ yields a non-local term. 

The imposition of this condition only on the anomaly term $a^0_0$, would still a priori allow a non-local solution to $b^0_0$  that canceled the non-localities of $\{H,\tilde b^{(2,3)}_0\}$, and thus a non-trivial local anomaly $a^0_0$ would still have been a strong logical possibility. However, we have shown that $\tilde{m}_0= m^0_0=b^0_0=0$ merely from the locality of the cohomology groups, so that this cannot occur even under such a  milder assumption of locality of the anomaly.  This demand would be softer than requiring all quantities $\tilde{m}_0,  m^0_0 , b^0_0, \dots $ to be local, but stronger than only requiring the gauge cohomologies $H^{(g,k)}(s_\omega | \tilde d)$ to be local.  Of course, since  the homogeneous solution $\bar a^0_0$ vanishes (since $H^{(1,3)}(s_{\mbox{\tiny W}} | \tilde d)=0$), and this is the only part of the anomaly that is independent of $H$, if we impose locality on  \emph{all} the solutions $\tilde{m}_0,  m^0_0 , b^0_0, \dots $, not just on the anomaly,  we would already obtain a trivial anomaly directly from \eqref{b0}, without actually needing to follow the descent equations. 

Thus we reach the ``middle-ground" result that if impose not only that the gauge cohomologies   $H^{(g,k)}(s_\omega | \tilde d)$ be local, but also the resulting total anomalies be local, then the vanishing of $\bar a^0_0$ and that of $\tilde{m}_0= m^0_0=b^0_0$ is sufficient to rule out the presence of anomalies. Note that of course the anomalous term coming from $[\Omega, \Omega]$  is already local, but its non-vanishing in the present circumstances requires a non-trivial solution $a^0_{0P}$, which indeed does vanish if we require locality.


\section{Conclusions}\label{sec:conclusions}

In this paper we have provided a first look into the issue of anomalies arising in the shape dynamics Weyl symmetry. 
Our strategy was to use cohomological methods for the special type of Weyl symmetry occurring in Shape Dynamics. 
We remind the reader that our Weyl symmetry is spatial, but acts on the metric \emph{as well as on its conjugate momenta}. Thus it is a type of phase space inherent symmetry that is not straightforwardly translatable to the space-time picture. 

For the study of Weyl anomalies we have used a Hamiltonian decomposition of the gauge cohomology, due to Barnich and Henneaux \cite{Barnich&Henneaux}. The most direct way to get to our result is to use this decomposition \eqref{equ:decomposition}:
\be H^{(g,k)}(s_H | d)\simeq H^{(g,k)}(s_\omega | \tilde d)\oplus l[H^{(g+1,k-1)}(s_\omega | \tilde d)]\oplus r[H^{(g,k-1)}(s_\omega | \tilde d)]\ee
where $l$ and $r$ are linear projections that depend  on the total Hamiltonian, $n-1$ is the maximal spatial dimension and $g$ is the ghost number (thus the anomaly would exist for $g=1, k=4$), and $\omega$ is the Hamiltonian BRST symmetry generator. By the above, the anomaly decomposes  into a purely spatial part, namely an anomaly in the symmetry itself, and a temporal part, respectively an anomaly in the time propagation of the symmetry.

For shape dynamics we show that  we have a decomposition of the symmetry into the diffeomrphism and Weyl part: $s_\omega=s_{\mbox{\tiny W}}+s_{\mbox{\tiny D}}$ similar to what happens in the usual Lagrangian setting, but in our case also acting of the momentum variables.  We then assume the Bardeem-Zumino shifting of the cohomology to the Weyl sector, and thus restrict our attention to the Weyl cohomology of spatially diffeomorphism (co)variant terms. The most important assumption of our work is that we consider only the local cohomology groups as the starting point to deriving our anomalies. Using this assumption,  we have shown that the first cohomology $H^{(1,n-1)}(s_{\mbox{\tiny W}}, \tilde d)$ of our spatial Weyl symmetry does \emph{not}  depend on the metric momenta, for arbitrary spatial dimensions. Thus we can use known results in the classification of the geometric anomalies \cite{Deser_anomalies},\cite{Bonora-Cotta-Ramusino, Bonora-Pasti-Bregola, Bonora-Pasti-Tonin} to calculate the cohomology spaces $H^{(1,n)}(s_{\mbox{\tiny W}}, \tilde d)$. 

For an odd dimensional space (such as in 3+1), there is no non-trivial ghost number one cohomology, which allows us to discard the $r[H^{(1,3)}(s_\omega | \tilde d)]$ component of the anomaly. This means that in the temporal anomaly 
$$ \int d^{n}x[\Omega, H]=\frac{\hbar}{2}\int  dt (a^0_{0P}+\bar{a}^0_0)
$$ what we termed the ``homogeneous" term $\bar{a}^0_0$ vanishes. To deal with the $a^0_{0P}$ part, we are required to investigate the ghost number 2 cohomology, $H^{(2,3)}(s_\omega | \tilde d)$.  Assuming that again the cohomology can be shifted to the Weyl sector and that it does not depend on the momenta, we focused on the Weyl sector of the cohomology $H^{(2,3)}(s_{\mbox{\tiny W}} | \tilde d)$ which has as its single basis element: 
$$ \tilde b^{(2,3)}_0= -2\sqrt{g}\eta \epsilon^{\alpha}_{\nu\rho\sigma}\eta_{,\alpha}(R^\nu_\mu -\frac{1}{4}\delta^\nu_\mu )dx^\mu dx^\rho dx^\sigma
$$
 We were then able to work out the modified descent equations  \eqref{equ:new_cocycle1}-\eqref{equ:new_cocycle2} and, using the shape dynamics equations of motion \eqref{equ:eom_g}-\eqref{equ:eom_pi}, to fully characterize the coefficients in the functional differential equation defining $a^0_{0P}$, which give:
 \be\label{equ:characterizing} s_{\mbox{\tiny W}} a^{0(1,3)}_{0P}+ \alpha\{H,\tilde b^{(2,3)}_0\}=0 
\ee
Due to the vanishing of $\bar{a}^0_0$, the temporal and spatial parts of the anomaly completely ``intertwine", and both will only exist if there is a non-trivial solution to equation \eqref{equ:characterizing}, in which case they are given by 
\begin{eqnarray}
\int d^{n}x[\Omega, \Omega]=\frac{\hbar}{2}\int \tilde b^{(2,3)}_{0}  dt  +\mathcal{O}(\hbar^2)\\
\int d^{n}x[\Omega, H]=\frac{\hbar}{2}\int a^{0(1,3)}_{0P} dt  +\mathcal{O}(\hbar^2)
\end{eqnarray}
So far we had only assumed locality of the cohomology groups $H^{(g,k)}(s_{\mbox{\tiny W}} | \tilde d)$, which lie at the basis for the anomaly calculations. However, if we furthermore assume locality of the anomaly itself, then there is no local solution to \eqref{equ:characterizing} and, although the spatial part of the anomaly is necessarily already local, the vanishing of $a^{0(1,3)}_{0P}$ requires both parts of  the anomaly, spatial and temporal,  to vanish. 


As a last comment, this work does not purport to be a final word in the issue of a cohomological treatment of anomalies in the context of shape dynamics. Our aim was to lay out some of the technical tools and obstacles in developing said treatment for shape dynamics, which we feel is a nice testing ground in which to employ the developments made for Hamiltonian anomalies in the late 90's. We should also stress that the issue of fully determining the temporal anomaly is a question that is likely to only be fully resolved with the 1-loop quantization of shape dynamics, which is work in progress. To put things in perspective, Loop Quantum Gravity also has not resolved the issue of anomalies in the Hamiltonian algebra. It would be interesting to see if these techniques apply also in that setting. 


\subsubsection*{Some issues with the solution.}

There are several  issues with the solution proposed here. Let us enumerate them: i) we have limited ourselves to looking for anomalies  formed from the \emph{local}  cohomology in the field space consisting only of local cochains in the phase space fields. This gives us a defining equation \eqref{equ:characterizing}. If we postulate furthermore that the anomaly itself is local, we can show that there are no solutions and thus no anomalies. This conclusion does not require that all   the ``particular" solutions $b^0_{0P}, m^0_{0P}, \dots$ be local, only that the last step leading to the anomaly is.   As we mentioned, if one allows arbitrary non-local cochains in the calculation of the Lagrangian anomaly, there are \emph{no} non-trivial cohomology elements to be found  in any theory \cite{Boulanger_general_WZ, Boulanger_algebraic_Weyl}.\footnote{Note for example that the Bardeem-Zumino non-local anomaly can be derived from a local anomaly \cite{Bertlmann}.} On the other hand, the Shape Dynamics Hamiltonian is non-local, and there might be some allowance of non-local terms in the space of cochains that does not trivialize the anomalies and yet is well-adapted to the Shape Dynamics ansatz. For example, one could allow the inverse Laplacian, or the York conformal factor to appear in  $a^0_{0P}$ ans see if with this extension one could obtain a non-trivial solution to \eqref{equ:characterizing}. It is our opinion that no such solution will exist. 

One could bypass this issue altogether and  attempt to look for such anomalies in the Linking Theory \cite{SD:LT}, but the search there is much harder because the equations \eqref{equ:counting_conf} and \eqref{equ:counting_tensor} do not limit the number of possible terms (basically any conformally transformed scalar $t_\phi F[g,\pi]$ qualifies). 

The second issue is that we have \emph{not} proven that one can shift the diffeomorphism anomalies to the Weyl anomalies by the addition of a Bardeem-Zumino counter-term to the action, as was done for the usual space-time anomalies by Bonora et al in \cite{Bonora-Cotta-Ramusino, Bonora-Pasti-Bregola, Bonora-Pasti-Tonin}. Although in principle this should go through rather straightforwardly for ghost number one,  one would need to prove it also for the second cohomology, as it is also required. 

There is still a third issue, which is that we have calculated the possible term coming only from the purely geometric sector (no dependence on the momenta). This was warranted for the Weyl ghost number one case as we can shift derivatives from the ghost scalars, and consider polynmials in the fields and fields momenta and their spatial derivatives as forming the cochain space. For ghost number two a more careful treatment is required, specially if one is not allowed to consider only the Weyl ghosts, as mentioned in the previous paragraph, since diffeomorphism ghosts carry a tensor index. As the intent of this paper is to provide an initial study into these questions, we leave the investigation of these issues to further study.

\section*{Acknowledgments}
We thank Charles-Melby Thompson for first discussing with us the Wess-Zumino condition. We thank Glenn Barnich for helpful clarifications, and we  specially thank Nicolas Boulanger for his hospitality and patience in carefully explaining the method utilized here, and UMONS for their hospitality. 
HG was supported in part by the U.S.
Department of Energy under grant DE-FG02-91ER40674.


\end{document}